\documentclass[aps,prl,showpacs,twocolumn,floats]{revtex4}
\usepackage{amssymb}

\usepackage{bm}

\begin{document}

\bibliographystyle{prsty}
\author{E. M. Chudnovsky}
\affiliation{Physics Department, Lehman College, The City
University of New York \\ 250 Bedford Park Boulevard West, Bronx,
New York 10468-1589, U.S.A.}
\date{11 May 2009}

\begin{abstract}
We compute correlation function of a superconducting order
parameter in a continuous model of a two-dimensional
Josephson-junction array in the presence of a weak Gaussian noise.
When the Josephson coupling is large compared to the charging
energy the correlations in the Euclidian space decay exponentially
at low temperatures regardless of the strength of the noise. We
interpret such a state as a collective Cooper-pair insulator and
argue that it resembles properties of disordered superconducting
films.
\end{abstract}

\title{Instanton Glass Generated by Noise in a Josephson-Junction Array}

\maketitle

Systems consisting of densely packed metallic grains have been
studied for decades, see, e.g., for review Ref.
\onlinecite{Beloborodov}. They exhibit peculiar electronic
properties that stem from the quantum tunneling of electrons
between the grains. Numerous models of Josephson-junction arrays
have been employed to describe properties of granular
superconductors \cite{Fazio}. The most recent theoretical research
in this area has been inspired by the experimental evidence of a
sharp low-temperature superconductor-insulator transition in
disordered films \cite{Shahar,Steiner,Baturina,Kowal} (see also
earlier experimental works \cite{old-exp}). Various theoretical
scenarios of the effect have been proposed: Suppression of the
superconducting order parameter by disorder in two dimensions
\cite{Finkel'stein}, Bose condensation of vortices
\cite{MPA-Fisher}, trapping of Cooper pairs due to Coulomb
blockade \cite{Fazio-Schon}, collective superinsulator phase
\cite{Vinokur}, insulating Josephson phase quenched by the
magnetic field \cite{Imry}, overheating of electrons due to
inefficient electron-phonon processes \cite{Altshuler1}, and
Anderson localization of Cooper pairs \cite{Altshuler2}.

In this Letter we are concerned with the correlations of the order
parameter in a two-dimensional array of strongly coupled
superconducting grains in the presence of a weak Gaussian noise.
We find that regardless of the strength of the Josephson coupling
between the grains, the imaginary-time correlation function of the
order parameter decays exponentially in 2+1 dimensions. We call
such a state {\it an instanton glass} and interpret it as a
collective Cooper-pair insulator in which Cooper pairs are weakly
localized within areas that include a large number of grains.

We consider superconducting grains that are sufficiently large so
that the fluctuations of the magnitude of the complex order
parameter $\Delta$ can be ignored. This condition is satisfied
when the distance between electron energy levels in the grain,
$\delta$, is small compared to $|\Delta|$. However, the phase
$\theta_i$ of $\Delta$ on each grain is a dynamical variable
described by the Hamiltonian
\begin{eqnarray}\label{ham}
{\cal{H}}  & = & \sum_i E_C^in_i^2
 + \sum_{\langle i<j \rangle}E_J^{ij}\left[1-\cos(\theta_i -
\theta_j)\right]
\nonumber \\
& + & \sum_{i}\epsilon_J^{i}\left[1-\cos(\theta_i -
\phi_i)\right]\,.
\end{eqnarray}
The sum is over all grains, with $\langle i<j\rangle$ denoting the
summation over the nearest-neighbor pairs of grains. The first
term in Eq.\ (\ref{ham}) corresponds to the charging effect due to
the Cooper-pair exchange between the grains, with $n_i$ being the
number operator for the excess Cooper pairs at the $i$-th grain
and $E_C^i$ being the charging energy of the grain.  The second
term describes the Josephson coupling of strength $E_J^{ij}$
between the grains. The form of the last term in Eq.\ (\ref{ham})
implies the existence of additional weak links that allow some
leakage of the Cooper pairs in and out of the superconducting
grains. In our model these weak links are external to the
two-dimensional Josephson-junction array of the grains. They exist
on top of the strong links between the grains. We assume random
distribution of phases $\phi_J$ and call such a disorder
``Josephson noise''.

The prevailing view is that the ground state of a granular
superconductor depends on the ratio of the Josephson coupling
energy and the charging energy. If this ratio is large, the Cooper
pairs move freely between the grains and the system is a
superconductor. If the ratio is small, which should be expected
for small grains with large $E_C = (2e)^2/(2C)$, then moving an
excess Cooper pair into the grain costs too much energy. In such a
case the Cooper pairs are localized on individual grains and the
system is an insulator. In our model we require
\begin{equation}\label{EJ}
E_J \gg E_C\,,
\end{equation}
so that our granular film is electrically close to a homogeneous
film. Quantitatively, this condition translates into a large
dimensionless tunneling conductance, $g = 2\pi\hbar/(e^2R) \ll 1$,
with $R$ being the normal resistance of the Josephson contact
between the grains. In this case the charging energy becomes
renormalized by the Coulomb screening of the excess charge on the
grain, so that \cite{Beloborodov}
\begin{equation}\label{EC}
E_C \approx \frac{|\Delta|}{g} \ll |\Delta| \,.
\end{equation}
Large $E_J$, provides small differences between $\theta_i$ and
$\theta_j$ at the neighboring grains. On the contrary, we assume
the Josephson noise to be weak,
\begin{equation}\label{weak-noise}
\langle\epsilon_J^2\rangle \ll \langle E_J^2\rangle\,,
\end{equation}
so that the corresponding small tunneling conductances, allow an
arbitrarily large difference between $\theta_i$ and $\phi_i$. In
what follows we treat $\phi_i$ as a dynamical random field. We
show that contrary to the previous findings, in the presence of
the noise, the Cooper pairs are localized at $T = 0$ within areas
of size  $r \propto 1/\langle\epsilon_J^2\rangle$ even under the
condition (\ref{EJ}).

Since $n_i$ and $\theta_i$ are canonically conjugated variables,
we have
\begin{eqnarray}
n_i & = & -i\frac{d}{d\theta_i} \\
 i\hbar\frac{d\theta_i}{dt} & = &
[\theta_i, {\cal{H}}] = 2iE_C^in_i\,.
\end{eqnarray}
This allows one to replace $n_i$ in Eq.\ (\ref{ham}) by
$\hbar{\dot{\theta}}_i/(2E_C^i)$. Then the first term in Eq.\
(\ref{ham}) acquires the form of the ``kinetic energy''. The
Euclidean action corresponding to the Hamiltonian (\ref{ham}) is
\begin{eqnarray}\label{action}
S_{eff}  & = & \int
d\tau\left\{\sum_i\frac{\hbar^2}{4E_C^i}\left(\frac{d{\theta}_i}{d\tau}\right)^2
\right.
\nonumber \\
& + & \sum_{\langle i<j\rangle}E_J^{ij}\left[1-\cos(\theta_i -
\theta_j)\right]
\nonumber \\
& + & \left.\sum_{i}\epsilon_J^{i}\left[1-\cos(\theta_i -
\phi_i)\right]\right\}\,,
\end{eqnarray}
where $\tau = it$. Without the kinetic term this action is
equivalent to the XY spin model in a random field that has been
intensively studied in the past \cite{XY}. Without the last term
Eq.\ (\ref{action}) has been also intensively studied (with and
without dissipation) in 1980s in connection with the possibility
of the low-temperature re-entrant superconductor - normal metal
transition due to quantum fluctuations of the phase
\cite{re-entrant}. Superconductor-insulator transition at $E_C
\sim 2E_J$ in a two-dimensional Josephson-junction array has been
confirmed by Monte-Carlo simulations \cite{Wallin}.  We are not
aware of any theoretical investigation of the ground state of the
model described by Eq.\ (\ref{action}) under the condition
(\ref{EJ}).

The essential features of the model can be studied by considering
a square Josephson-junction array with a lattice spacing $a$,
$E_C^i = E_C$, $E_J^{ij}= E_J$ at $T = 0$. Small difference of the
phase for the neighboring grains allows one to write for the
nearest neighbors
\begin{eqnarray}
& & \cos[\theta({\bf r_i}) - \theta({\bf r}_j)] \, \approx \, 1
-\frac{1}{2}[\theta({\bf r_i}) - \theta({\bf r}_j)]^2 \\
& & \theta({\bf r}_j)\, \approx\, \theta({\bf r_i}) + ({\bf r}_j -
{\bf r}_i)\cdot [{\bm \nabla}\theta({\bf r})]_{{\bf r} = {\bf
r}_i}\,.
\end{eqnarray}
Substitution of these equations into Eq.\ (\ref{action}),
summation over the four nearest neighbors in the square lattice,
and replacement of the summation over $i$ by the integration
according to $\sum_i \rightarrow \int d^2r/a^2$, yields a
continuous field model described by the action
\begin{eqnarray}\label{continuous}
& & S_{eff}  =  \int d\tau \int d^2r
\left\{\frac{\hbar^2}{4a^2E_C}\left(\frac{d\theta}{d\tau}\right)^2
 + E_J \left(\frac{d\theta}{d{\bf r}}\right)^2 \right\} \nonumber  \\
& & +    \int d\tau \int \frac{d^2r}{a^2}\, \epsilon_J({\bf r})
\left\{1-\cos[\theta({\bf r,\tau}) - \phi({\bf r},\tau)]\right\}
\,.
\end{eqnarray}

It is convenient to use dimensionless variables:
\begin{equation}
\bar{x} = \frac{x}{a}\,, \quad \bar{y} = \frac{y}{a}\,, \quad
\bar{\tau} = \frac{\tau}{{\tau}_0}
\end{equation}
with
\begin{equation}
\tau_0 = \frac{\hbar}{2\sqrt{E_JE_C}}\,.
\end{equation}
In terms of these variables Eq.\ (\ref{continuous}) acquires a
simple form:
\begin{eqnarray}\label{dim-action}
\frac{S_{eff}}{\hbar} & = & \left(\frac{E_J}{E_C}\right)^{1/2}\int
d^3\bar{r}\left\{\frac{1}{2}(\bar{\bm \nabla}\theta)^2
\right.\nonumber
\\
& + & \left.\frac{\epsilon_J}{2E_J}\left[1 - \cos(\theta -
\phi)\right]\right\}.
\end{eqnarray}
Here the integration is over dimensionless Euclidian coordinates
($d^3r = d\bar{x}d\bar{y}d\bar{\tau}$), $\bar{\bm \nabla}\theta$
is the 3d gradient of $\theta$ with respect to these coordinates.

We are interested in the limit of $E_J \gg E_C$ when ${S_{eff}}$
is large compared to ${\hbar}$ and the phase $\theta(\bar{\bf r})
$ is a well-defined semiclassical field. Quantum dynamics of such
a field is dominated by the extremal trajectories of Eq.\
(\ref{dim-action}) satisfying
\begin{equation}\label{extremal}
\bar{\bm \nabla}^2 \theta = \frac{\epsilon_J}{2E_J} \sin(\theta -
\phi) \,.
\end{equation}
At $\epsilon_J = 0$ this equation possesses a solution $\bar{\bm
\nabla} \theta = {\rm const}$ that describes a global
superconducting current. In general, for such a current to exist,
the phases at distant points must be correlated. We, therefore,
want to compute the correlation function
\begin{eqnarray}\label{correlator}
& & C(\bar{\bf r}_1,\bar{\bf r}_2) \equiv
\frac{\langle\Delta(\bar{\bf r}_1)\Delta^{\dagger}(\bar{\bf
r}_2)\rangle}{|\Delta|^2} =
  \nonumber \\
& & \langle e^{i[\theta(\bar{\bf r}_1) - \theta(\bar{\bf
r}_2)]}\rangle =   \langle \cos[\theta(\bar{\bf r}_1) -
\theta(\bar{\bf r}_2)]\rangle\,,
\end{eqnarray}
where ${\bf r} = (\bar{x},\bar{y},\bar{\tau})$. The average in
Eq.\ (\ref{correlator}) is over all pairs of points $(\bar{\bf
r}_1, \bar{\bf r}_2)$ in 2+1 dimensions that are separated by the
same distance $|\bar{\bf r}_1 - \bar{\bf r}_2|$.

The non-linear dynamics of the field expressed by Eq.\
(\ref{extremal}) usually presents a problem for the computation of
the correlation function in Eq.\ (\ref{correlator}). Below we use
a mathematical trick that under the condition (\ref{weak-noise})
allows one to obtain $C(\bar{\bf r}_1,\bar{\bf r}_2)$ exactly with
a conventional choice for the noise. We introduce a random
two-component vector field,
\begin{equation}
{\bf f}(\bar{\bf r}) = [f_1,f_2] = [\epsilon_J(\bar{\bf
r})\cos\phi(\bar{\bf r}), \epsilon_J(\bar{\bf r})\sin\phi(\bar{\bf
r})]\,,
\end{equation}
and write Eq.\ (\ref{extremal}) in the integral form:
\begin{eqnarray}\label{integral}
& & \theta(\bar{\bf r}) = \frac{1}{2E_J}\int d^3\bar{r}'G(\bar{\bf
r} - \bar{\bf r}' ) \times \nonumber \\
& & [f_1(\bar{\bf r}')\sin\theta(\bar{\bf r}') - f_2(\bar{\bf
r}')\cos\theta(\bar{\bf r}')]\,,
\end{eqnarray}
where
\begin{equation}
G(\bar{\bf r})= -\frac{1}{4 \pi|\bar{\bf r}|}
\end{equation}
is the Green function of the 3d Laplace equation, satisfying
$\bar{\bm \nabla}^2G(\bar{\bf r}) = \delta(\bar{\bf r})$.
Substituting Eq.\ (\ref{integral}) into Eq.\ (\ref{correlator})
one obtains
\begin{eqnarray}\label{cor-int}
& & C(\bar{\bf r}_1,\bar{\bf r}_2) = \langle
\exp\left\{\frac{i}{2E_J} \int d^3\bar{r} [G(\bar{\bf r}_1 -
\bar{\bf r} ) - G(\bar{\bf r}_2 - \bar{\bf r} )]\right.
\nonumber \\
& & \times \left.[f_1(\bar{\bf r})\sin\theta(\bar{\bf r}) -
f_2(\bar{\bf r})\cos\theta(\bar{\bf r})]\right\}\rangle \,,
\end{eqnarray}

To proceed with the calculation of the space average in Eq.\
(\ref{cor-int}) one needs to choose the model of the noise. The
simplest choice corresponds to the Gaussian distribution for the
probability, $P$, of any given realization ${\bf f}(\bar{\bf r})$:
\begin{equation}
P[{\bf f}(\bar{\bf r})] \propto \exp\left[-\frac{1}{2\langle
\epsilon_J^2\rangle}\int d^3\bar{r}\,{\bf f}^2(\bar{\bf
r})\right]\,,
\end{equation}
which also provides definition of $\langle \epsilon_J^2\rangle$.
With this assumption Eq.\ (\ref{cor-int}) becomes
\begin{eqnarray}\label{disorder-int}
& & C(\bar{\bf r}_1,\bar{\bf r}_2) = \left[\int D^2\{{\bf
f}\}\exp\left\{-\frac{1}{2\langle \epsilon_J^2\rangle}\int
d^3\bar{r}\,{\bf
f}^2\right\}\right]^{-1} \times \nonumber \\
& &  \int D^2\{{\bf f}\} \exp\left\{\int d^3\bar{r}
\left(\frac{i}{2E_J}[G(\bar{\bf r}_1 - \bar{\bf r} ) - G(\bar{\bf
r}_2 - \bar{\bf r} )]  \right.\right.
\nonumber \\
& & \times \left. \left.[f_1(\bar{\bf r})\sin\theta(\bar{\bf r}) -
f_2(\bar{\bf r})\cos\theta(\bar{\bf r})] - \frac{{\bf
f}^2}{2\langle \epsilon_J^2\rangle}\right)\right\}\,,
\end{eqnarray}
where $\int  D^2\{{\bf f}(\bar{\bf r})\}= \int D\{f_1\}D\{f_2\}$
denotes functional integration over the realizations of disorder.

At first glance the evaluation of the above correlator may seem
hopeless because it requires the explicit knowledge of
$\theta(\bar{\bf r})$ created by ${\bf f}(\bar{\bf r})$. To see
that $C(\bar{\bf r}_1,\bar{\bf r}_2)$ can be calculated exactly in
the limit of weak noise, we first notice that according to Eq.
(\ref{weak-noise}) and Eq.\ (\ref{extremal}) the contribution of
${\bf f}(\bar{\bf r})$ to the spatial derivatives of
$\theta(\bar{\bf r})$ is generally small. In fact, as is shown
below, the significant change in $\theta(\bar{\bf r})$ occurs over
the distances $\bar{r} \sim E_J^2/\langle\epsilon_J^2\rangle \gg
1$. This means that the value of $\theta$ at a certain point
$\bar{\bf r}$ has very little correlation with  ${\bf f}$ at that
point. Consequently, to the lowest order on the noise, the
dependence of $\theta(\bar{\bf r})$ on ${\bf f}(\bar{\bf r})$ in
Eq.\ (\ref{disorder-int}) can be neglected and the remaining
Gaussian integration over ${\bf f}$ can be easily performed. As a
result, $\sin\theta(\bar{\bf r})$ and $\cos\theta(\bar{\bf r})$ in
the exponent nicely combine into $\sin^2\theta(\bar{\bf r}) +
\cos^2\theta(\bar{\bf r})=1$, yielding
\begin{eqnarray}
& & C  =
\exp{\left\{-\frac{\langle\epsilon_J^2\rangle}{8E_J^2}\int
d^3\bar{r} [G(\bar{\bf r}_1 - \bar{\bf r} ) -
G(\bar{\bf r}_2 - \bar{\bf r} )]^2\right\}} \nonumber \\
& & =  \exp\left\{-\frac{\langle\epsilon_J^2\rangle}{4E_J^2}\int
\frac{d^3\bar{q}}{(2\pi)^3} \frac{1 - \cos[\bar{\bf
q}\cdot(\bar{\bf r}_1 - \bar{\bf r}_2
)]}{\bar{q}^4}\right\}.\nonumber \\
\end{eqnarray}
Further integration gives
\begin{equation}\label{cor-final}
C(\bar{\bf r}_1,\bar{\bf r}_2) = \exp{\left(-\frac{|\bar{\bf r}_1
- \bar{\bf r}_2|}{l}\right)}\,, \qquad l = \frac{32\pi
E_J^2}{\langle\epsilon_J^2\rangle}\,.
\end{equation}

Phase correlations that decay exponentially in both space and
imaginary time indicate that the Josephson-junction array is in
the insulating state \cite{Sondhi}. Since the instanton solutions
of Eq.\ (\ref{extremal}) describe tunneling trajectories of Cooper
pairs, we call such a state an ``instanton glass''. The presence
of a finite correlation length in the $(x,y)$ plane,
\begin{equation}\label{R}
R_c = \frac{32 \pi E_J^2}{\langle\epsilon_J^2\rangle}\,a\,,
\end{equation}
implies that at any given moment of time the phase correlation is
lost over spatial distances greater than $R_c$. At any point in
space there is also a finite correlation length in the imaginary
time,
\begin{equation}
\tau_c = \frac{32 \pi
E_J^2}{\langle\epsilon_J^2\rangle}\,\tau_0\,,
\end{equation}
that implies the existence of the energy gap,
\begin{equation}\label{gap}
\Delta_{IG} = \frac{\hbar}{\tau_c} =
\frac{\langle\epsilon_J^2\rangle}{16\pi
E_J}\left(\frac{E_C}{E_J}\right)^{1/2}\,,
\end{equation}
characteristic of an insulator \cite{Sondhi}. It represents the
localization energy of Cooper pairs within overlapping areas of
size $R_c$.

Applied to a granular film, the above results mean that Cooper
pairs are localized within regions of size $R_c$ that are large
compared to the average size of the grain $a$. If the dimensions
of the film, $L$, are greater than $R_c$, the film should be a
Cooper-pair insulator. At $T \ll \Delta_{IG}$ the conductivity of
such a film must be due to the thermal hopping of Cooper pairs
between regions of size $R_c$, obeying the law
$\exp(-\Delta_{IG}/T)$. So far we have not included the magnetic
field into the problem. Its treatment within our model is much
more involved but a plausible speculation can be made about the
expected effect of the field. The latter is known to suppress
Josephson tunneling. The weaker the coupling the smaller is the
critical field that destroys tunneling. In our model the weakest
links are the external ones characterized by the coupling strength
$\epsilon_J$. They will be destroyed by the field  first, making
the localization length (\ref{R}) infinite and destroying the
insulating gap (\ref{gap}). The non-monotonic dependence of the
Josephson coupling on the field \cite{Tinkham} may lead to the
non-monotonic field dependence of the gap. The expected
temperature and field behavior of the instanton glass is,
therefore, in a qualitative agreement with experimental findings
in disordered superconducting films \cite{Baturina}.

Also in agreement with experiment is the fact that such a behavior
is pertinent to a two-dimensional system and does not appear in a
three-dimensional superconducting sample. Indeed, for a 3d sample,
the calculation similar to the one performed above, but with the
Green function $G_4(\bar{\bf r})= -1/(4 \pi^2|\bar{\bf r}|^2)$
instead of $G_3(\bar{\bf r})= -1/(4 \pi|\bar{\bf r}|)$, provides a
power-law decay of the correlations of the phase in a 3+1
Euclidian space. The corresponding 3+1 correlation length is,
therefore, infinite and the gap is zero. Note that such a
dependence of the correlations on the dimensionality of space is
typical for models with a continuous order parameter in the
presence of quenched disorder \cite{Larkin,Imry-Ma}. Our approach
is, in effect, an extension of the Larkin-Imry-Ma theorem to the
Euclidean space-time for problems that involve tunneling in the
presence of noise.

In conclusion, we have demonstrated that Josephson noise, no
matter how weak, destroys long-range space-time correlations of
the order parameter in a two-dimensional Josephson-junction array
regardless of the tunneling conductance. We have argued that this
effect can be relevant to the observed low-temperature
superconductor-insulator transition in disordered films.

The author acknowledges illuminating discussions with D. Kabat, A.
Kuklov, and V. Vinokur. This work has been supported by the
Department of Energy through Grant No. DE-FG02-93ER45487.


\begin{thebibliography}{99}

\bibitem{Beloborodov}
See for review I. S. Beloborodov, A. V. Lopatin, V. M. Vinokur,
and K. B. Efetov, Rev. Mod. Phys. {\bf 79}, 469 (2007).

\bibitem{Fazio}
See for review R. Fazio and H. S. J. van der Zant, Phys. Rep. {\bf
355}, 235 (2001).

\bibitem{Shahar}
G. Sambandamurthy, L. W. Engel, A. Johansson, and D. Shahar, Phys.
Rev. Lett. {\bf 92}, 107005 (2004); G. Sambandamurthy, L. W.
Engel, A. Johansson, E. Peled, and D. Shahar, Phys. Rev. Lett.
{\bf 94}, 017003 (2005); M. Ovadia, B. Sac\'{e}p\'{e}, and D.
Shahar, Phys. Rev. Lett. {\bf 102}, 176802 (2009).

\bibitem{Steiner}
M. A. Steiner and A. Kapitulnik, Physica C {\bf 422}, 16 (2005);
M. A. Steiner, N. P. Breznay, and A. Kapitulnik, Phys. Rev. B {\bf
77}, 212501 (2008).

\bibitem{Baturina}
T. I. Baturina, C. Strunk, M. R. Baklanov, and A. Satta, Phys.
Rev. Lett. {\bf 98}, 127003 (2007); T. I Baturina, A. Yu. Mironov,
V. M. Vinokur, M. R. Baklanov, and C. Strunk, Phys. Rev. Lett.
{\bf 99}, 257003 (2007).

\bibitem{Kowal}
D. Kowal and Z. Ovadyahu, Physica C {\bf 468}, 322 (2008).

\bibitem{old-exp}
M. Strongin, R. S. Thompson, O. F. Kammerer, and J. E. Crow, Phys.
Rev. B {\bf 1}, 1078 (1970); D. B. Haviland, Y. Liu, and A. M.
Goldman, Phys. Rev. Lett. {\bf 62}, 2180 (1989); S. J. Lee and J.
B. Ketterson, Phys. Rev. Lett. {\bf 64}, 3078 (1990); D. Shahar
and Z. Ovadyahu, Phys. Rev. B {\bf 46}, 10917 (1992); H. S. J. van
der Zant, F. C. Fritschy, W. J. Elion, L. J. Geerligs, and J. E.
Mooij, Phys. Rev. Lett. {\bf 69}, 2971 (1992); Y. Liu, D. B.
Haviland, B. Nease, and A. M. Goldman, Phys. Rev. B {\bf 47}, 5931
(1993); A. Yazdani and A. Kapitulnik, Phys. Rev. Lett. {\bf 74},
3037 (1995); F. Ladieu, M. Sanquer, and J. P. Bouchaud, Phys. Rev.
B {\bf 53}, 973 (1996); N. Mason and A. Kapitulnik, Phys. Rev.
Lett. {\bf 82}, 5341 (1999).

\bibitem{Finkel'stein}
A. M. Finkel'stein, Pis'ma Zh. Eksp. Teor. Fiz. {\bf 45}, 37
(1987) [Sov. Phys. JETP Lett. {\bf 45}, 46 (1987)].

\bibitem{MPA-Fisher}
M. P. A. Fisher, Phys. Rev. Lett. {\bf 65}, 923 (1990).

\bibitem{Fazio-Schon}
R. Fazio and G. Sch\"{o}n, Phys. Rev. B {\bf 43}, 5307 (1991).

\bibitem{Vinokur}
V. M. Vinokur, T. I. Baturina, M. V. Fistul, A. Yu. Mironov, M. R.
Baklanov, and C. Strunk, Nature {\bf 452}, 613 (2008); M. V.
Fistul, V. M. Vinokur, and T. I. Baturina, Phys. Rev. Lett. {\bf
100}, 086805 (2008), see a more detailed model in arXiv:0806.4311.

\bibitem{Imry}
Y. Imry, M. Strongin, and C. C. Homes, Physica C {\bf 468}, 288
(2008).

\bibitem{Altshuler1}
B. L. Altshuler, V. E. Kravtsov, I. V. Lerner, and I. L. Aleiner,
Phys. Rev. Lett. {\bf 102}, 176803 (2009).

\bibitem{Altshuler2}
S. V. Syzranov, K. B. Efetov, and B. L. Altshuler,
arXiv:0903.3610.

\bibitem{XY}
A. Aharony and E. Pytte, Phys. Rev. Lett. {\bf 45}, 1583 (1980);
A. Houghton, R. D. Kenway, and S. C. Ying, Phys. Rev. B {\bf 23},
298 (1981); J. L. Cardy and S. Ostlund, Phys. Rev. B {\bf 25},
6899 (1982); E. M. Chudnovsky and R. A. Serota, Phys. Rev. B {\bf
26}, 2697 (1982); E. M. Chudnovsky, W. M. Saslow, and R. A.
Serota, Phys. Rev. B {\bf 33}, 251 (1986); P. Le Doussal and T.
Giamarchi, Phys. Rev. Lett. {\bf 74}, 606 (1995).

\bibitem{re-entrant}
D. M. Wood and D. Stroud, Phys. Rev. B {\bf 25}, 1600 (1982); L.
Jacobs, J. V. Jos\'{e}, and M. A. Novotny, Phys. Rev. Lett. {\bf
53}, 2177 (1984); S. Chakravarty, G.-L. Ingold, S. Kivelson, and
A. Luther, Phys. Rev. Lett. {\bf 56}, 2303 (1986).

\bibitem{Wallin}
M. Wallin, E. S. S{\o}rensen, S. M. Girvin, A. P. Young, Phys.
Rev. B {\bf 49}, 12115 (1994).

\bibitem{Sondhi}
S. L. Sondhi, S. M. Girvin, J. P. Carini, and S. Shahar, Rev. Mod.
Phys. {\bf 69}, 315 (1997).

\bibitem{Tinkham}
M. Tinkham, D. W. Abraham, and C. J. Lobb, Phys. Rev. B {\bf 28},
6578 (1983).

\bibitem{Larkin}
A. I. Larkin, Zh. Eksp. Teor. Fiz. {\bf 58}, 1466 (1970) [Sov.
Phys. JETP {\bf 31}, 784 (1970)]; A. I. Larkin and Yu. M.
Ovchinnikov, J. Low Temp. Phys. {\bf 34}, 409 (1979).

\bibitem{Imry-Ma}
Y. Imry and S. Ma, Phys. Rev. Lett. {\bf 35}, 1399 (1975).

\end{thebibliography}
\end{document}